\documentclass[12pt]{article}

\newcommand{\be}{\begin{equation}}
\newcommand{\ee}{\end{equation}}

\newcommand{\bea}{\begin{eqnarray}}
\newcommand{\eea}{\end{eqnarray}}
\begin{document}
\title{The universal $\psi''/\psi'$ ratio as unambiguous signature for
hard physics in nuclear reactions and in decays of beauty hadrons}
\author{
L.~Frankfurt${}^a$, L.~Gerland${}^a$, M.~Strikman${}^b$, M.~Zhalov${}^c$ \\
{\small\it ${}^a$School of Physics and Astronomy, Raymond and Beverly 
Sackler }\\
{\small \it
Faculty of Exact Science, Tel Aviv University, Ramat Aviv 69978,
Tel Aviv , Israel}\\
{\small\it ${}^b$Pennsylvania State University, University Park, 
Pennsylvania 16802}\\
{\small\it ${}^c$Petersburg Nuclear Physics Institute, Gatchina 188350, 
Russia}
} 
\maketitle
\begin{abstract}

We suggest the ratio of the $\psi''(3770)/\psi'(3686)$ production cross
sections as an unambiguous probe of the dominance of hard, point-like
physics in charmonium production.  If hard physics dominates the charmonium
production the ratio : ${\sigma(\psi'')\over\sigma(\psi')}\approx
{\Gamma(\psi''\to l^+l^-)\over \Gamma(\psi'\to l^+l^-)} \approx 0.1$ is
independent of the process. We argue that since the dominant $D\bar D$ decay
mode of the $\psi''$-meson can be detected with a high efficiency at the
current facilities the effective counting rate for the $\psi''$ production
should be comparable to that of the $\psi'$ and hence it could be observed
at the $B$-factories.  Even in hadron induced processes the ratio will not
be changed significantly by final state interactions due to the similar mass
and size of these two charmonium states. A possible exception are the
charmonium state interactions with low energy particles -- comovers -- which
are especially relevant for the heavy ion collisions.

\end{abstract}

The production of the charmonium states in inclusive hadron-hadron
collisions remains one of the challenges for the theory. Understanding of
the charmonium production dynamics is important for a number of
applications.  This includes the dynamics of $ B\to J/\psi + X$ decays which
play an important role in the investigations of the CP violation, see e.g. 
\cite{babar,cleo}. The $J/\psi$ production is also often considered as an
effective probe of the heavy ion collisions.  The matter is further
complicated by the question whether the mass of the $c$-quark is
sufficiently large as compared to the scale of the strong interaction to
apply perturbation theory for the production of charm quark pairs in pp, pA,
and AA collisions as well as in the decay of $B$-mesons.

So far experimental studies of charmonium states focused on the production
of $J/\psi,\,\psi',$ and $\chi$-states. The aim of this letter is to draw
attention to the production of the $\psi''$ ($\psi(3770)$) meson as a method
to probe the hardness of the process.  Really in hard processes like in the
process:  $e\bar e\rightarrow$~charmonium~state the contribution of S states
should dominate the $c\bar c$ systems independently whether S or D states
dominate in the wave function of charmonium state $\phi(k)$ in an average
charmonium configuration. This is because the matrix element of the process
is $\propto \int \phi(k)d^3 k \propto \phi(r=0)$. But in a nonrelativistic
charmonium model the wave function of a state with orbital momentum L e.g.\
the D wave is $\propto r^L$ at small relative interquark distances r because
of the orbital momentum barrier. Thus the ratio of $\psi'$ to $\psi''$
($\psi(3770)$) should be the same for any process of hard production of
charmonium states.  The universality will be lost for the processes where
the production will be influenced by nonperturbative QCD physics. So in this
paper, we compare our results with the predictions of the nonrelativistic
QCD (NRQCD) approach to the production of the onium
states~\cite{Bodwin:1994jh} which allowed to describe several features of
the data. (Note that so far this approach meets problems in a number of
cases - see ref.~\cite{Bodwin:2002mr} for a recent review.) The NRQCD
approach of ref.~\cite{nrqcd} assumed the dominance of D waves in the
produced $c\bar c$ quark system which means nonhard production mechanisms
play an important role. (The authors claims that the uncertainties in the
evaluation of the relevant amplitudes may be one order of magnitude.) So we
compare predictions of the NRQCD approach of ref.~\cite{nrqcd} and hard
production mechanisms for the BELLE and the BABAR experiments. 

We argue that the experimental observation of this process is feasible at
the current facilities (which have a very good efficiency for detection of
$D-$mesons), and would allow to check whether the $\psi'$ and $\psi''$ are
produced by a $c\bar c$ at short distances\footnote{An important advantage
of comparing $\psi''$ and $\psi'$ production as compared to the case of
$J/\psi$ and $\psi'$ is a near degeneracy of the masses of these states but
different dominating decay channels.}.  Our suggestion is based on the well
known observation that the properties of the $\psi''$ are well described
(actually has been predicted) using the charmonium model by
Eichten~\cite{Eichten:1974af,Richard:1979fc}.  Within this model it is found
that the $\psi''$ is the $1D$ state with a small admixture of the $2S$-wave
and correspondingly $\psi'$ is the $2S$ state with a small admixture of the
$1D$-wave. Namely
\begin{eqnarray}
\left|\psi'\right\rangle&=&\cos \theta \left|2S\right\rangle + \sin\theta
\left|1D\right\rangle\quad,\cr \left|\psi''\right\rangle&=&\cos \theta
\left|1D\right\rangle - \sin\theta \left|2S\right\rangle\quad. 
\label{mixing} 
\end{eqnarray} 
Since only the $S$-wave contributes to the decay of $\psi$ states into
$e^+e^-$-pairs (at least in nonrelativistic charmonium models) the value of
$\theta=19\pm 2^o $ can be determined from the data on the $e^+e^-$ decay
widths of $\psi'$ and $\psi''$ as\footnote{An estimate of
ref.~\cite{Richard:1979fc} within the charmonium model indicate that due to
recoil corrections of the electromagnetic current the D-wave gives a $\sim
20\%$ contribution to the e.m.  decay width. However, this model with the
same parameters yields a too large electromagnetic decay width for the
S-state.} 
\begin{equation} 
\tan^{2}(\theta)= {\Gamma (\psi''\to e^+ e^-)\over \Gamma(\psi'\to e^+
e^-)}\quad . 
\label{3} 
\end{equation} 
If the charmonium production is a short distance process, then it is
proportional to the square of the wave function of charmonium at zero
distance. Therefore, only the $S$-wave can contribute and the $\psi''/\psi'$
ratio is universal, i.e.\ not energy dependent and not dependent on the
process.  Since the masses of $\psi''$ and $\psi'$ are practically the same,
the difference of the wave functions in the origin is the only source of the
difference of the production rates. Correspondingly, the mixing model allows
us to predict
\begin{equation}
{\sigma(\psi'')\over\sigma(\psi')}=\tan^2 (\theta) \approx 0.1\quad. 
\label{ratiohard} 
\end{equation}

It is instructive to consider the produced $c\bar c$ as a coherent
superposition of the hadronic states.  Since $m_{\psi'}\approx m_{\psi''}$
the coherence of the $c\bar c$ configuration which ultimately transforms
into either $\psi'$ or $\psi''$ holds up to the large distances $l_c \sim
2P/(m_{\psi''}^2 -m_{\psi'}^2)\approx \gamma\cdot 2.3fm$, where $\gamma$ is
the $\gamma$-factor of the meson. $l_c$ exceeds the size of the interaction
region for all conceivable kinematics except deeply in the nucleus
fragmentation region.  Therefore it is more appropriate to think of the
propagation of the $2S$-state through the media with a subsequent
decomposition of this state into $\psi'$ and $\psi''$. 

The hadronic final state interactions cannot change significantly the
predicted ratio. Indeed, soft QCD interactions cannot transform the
$S$-state to $D$-state with significant probability. Significant values for
the cross sections of the nondiagonal exclusive diffractive processes are a
signature in the hadronic basis for the color screening phenomenon in the
interaction of small color singlets~\cite{FS91} which should be a correction
in soft QCD processes. The same conclusion is valid in the PQCD model for
the charm dipole nucleon interactions. Hence the prediction of eq.~(\ref{3})
appears to be pretty robust provided $\psi$ are produced via hard 
process\footnote{Note that though the mass resolution in the current
experiments is
not good enough to resolve $\psi'$ and $\psi''$ in the lepton mode the
contribution of $\psi''$ into such decays is negligible due to much smaller
value of $\Gamma_{e^+ e^-}/\Gamma_{tot}$.}. At the same time if soft
dynamics determines production of $\psi'',\psi'$ a stronger suppression of
the $\psi''$ production compared to the $\psi'$ production seems to be not
natural since the charmonium models lead practically to the same sizes of
the two mesons. For example, the model of ref.~\cite{Eichten:1974af} gives
for the average interquark distances $r_{2S}=0.760$~fm and
$r_{1D}=0.786$~fm. 

Let us consider the consequences of the prediction given by
eq.~(\ref{ratiohard}) for the $B$-meson decays now.  The branching fraction
for B-decays into the $\psi(2S)$ was found by the BABAR-collaboration to be
$0.297\pm0.020\pm0.020\%$~\cite{babar}.  This branching fraction is averaged
over the different B-mesons. The errors are statistical and systematical,
respectively.  The CLEO-collaboration found a comparable value of
$0.316\pm0.014\pm0.023\pm0.016\%$~\cite{cleo}. The third error is an overall
scale factor. Due to the chosen decay channels of the $\psi(2S)$
($\psi(2S)\to $~dileptons and $\psi(2S)\to J/\psi+\pi^++\pi^-)$ we expect
that these mesons are mostly $\psi'$ states with a negligible admixture of
$\psi''$ states. With the ratio of eq.~(\ref{ratiohard}) we predict the
branching fractions for decays of the $B$-mesons into the $\psi''$ to be
approximately $0.03\%$. In contrast the NRQCD approach predicts
$B(B\to\delta_1+X)=0.28\%$~\cite{nrqcd}, where $\delta_1$ is the D-wave
charmonium with the quantum numbers of the $\psi''$.

This branching fractions for decays of the $B$-mesons into the $\psi''$
seems to be smaller than the typical errors of these experiments. However,
the statistics may not be problematic since in the measurement of the $B\to
\psi(2S)$ both the BABAR and CLEO used the dileptonic decay channels of the
$\psi(2S)$ or of the $J/\psi$ which are smaller than $1\%$ and $10\%$,
respectively, while the decay of the $\psi''$ into $D$-mesons is the
dominating channel.  For this channel the detection efficiency is much
higher. Hence it seems likely that the counting rate for production of
$\psi''$ via $D\bar D $ decays could be comparable to that for the $\psi'$
decays via dilepton channels.  If so this could provide a cross check of the
measurement of the CP violation using new channels involving $\psi''$
production. In the refs.~\cite{cleo2,belle} the collaborations CLEO and
Belle showed that the channel $B^{\pm,0}\to \psi(2S)+K^{\pm,0}$ is 20\% as
large as the same channel for the $J/\psi$. However, due to the chosen decay
channels this is the contribution of the $\psi'$ only and the corresponding
$\psi''$ may add another 20\% if measured via the $D\bar D $ decays.

Additionally, the Belle Collaboration reported recently about the double
charmonium production in $e^ +e^-$ annihilation at
$10.6$~GeV~\cite{belle2}\footnote{We are grateful to A.\ Dumitru for drawing
our attention to the refs.~\cite{belle,bodwin} and for discussions.}. In
different theoretical works (one of the most recent is ref.~\cite{bodwin})
is discussed if the cross section for double charmonium production can be
described within a perturbative framework. However, the ratio of
eq.~(\ref{ratiohard}) may be used here to test if these charmonium states
result from local processes ($e^+ e^-$ annihilation, or PQCD) or from soft
processes in the jet fragmentation. 
The BELLE collaboration found a production cross section for the $\psi'$ of
$\sigma(e^+e^-\to\psi'+X)=0.67\pm0.09$~pb. The cross section of
the $\psi''$ calculated in ref.~\cite{nrqcd} with the NRQCD approach is
$\sigma(e^+e^-\to\delta_1+X)=0.043$~pb.
This means that they predict roughly a factor of two more $\psi''$
than we, since this D-wave channel has the same order as the
contribution of the S-wave. 

In $p\bar p$ collisions at the Tevatron the CDF-collaboration found an
inclusive production cross section for the $\psi'$ of $\sigma(p\bar p\to
\psi'+X)=30.1\pm 6.6^{+3.8}_{-4.2}\pm 6.6$~nb~\cite{cdf}. The errors are
statistical, systematical, and due to the uncertainty of the
$\psi'\to\mu^+\mu^-$ branching ratio\footnote{With the latest value for the
$\psi'\to\mu^+\mu^-$ branching ratio of ${\Gamma(\psi'\to
\mu^+\mu^-)\over\Gamma(\psi'\to X)}=0.007\pm 0.0009$~\cite{pdg} this would
be $\sigma(p\bar p\to \psi'+X)=33.1\pm 6.6^{+3.8}_{-4.2}\pm 4.3$~nb.}. Since
the $\psi'$ is measured here via the decay into $\mu^+\mu^-$ pairs, there is
nearly no admixture of the $\psi''$. Therefore we predict with
eq.~(\ref{ratiohard}) an inclusive production cross section for the $\psi''$
of $\sigma(p\bar p\to \psi''+X)\approx 3$~nb.

The $\psi'$ meson production was also studied with 800 GeV protons at the
Fermilab by the E789 and E772 collaboration using a set of nuclear
targets~\cite{e789,e772}. These data together with eq.~(\ref{ratiohard})
yield
for the integrated over $x_F$ and $p_t$ production cross section of the
$\psi''$ on a hydrogen target
\begin{equation}
\sigma(pp\to \psi''+X)=40\pm 19\mbox{ nb}\quad.
\end{equation}
Charmonium is also measured in proton nucleus collisions at a slightly
higher energy ($E_{lab}=920$~GeV) at HERA-B~\cite{herab}.

Note that several recent analyses of charmonium production processes
performed within the NRQCD approach, see e.g.~\cite{nrqcd}, have suggested
the possibility of a significant production cross section of D-wave states. 
We compared our results for the BABAR and the BELLE experiment with the
NRQCD predictions. According to NRQCD in these processes the direct
contribution to the D-wave charmonium states is as large as the
contribution to the $\psi''$ due to the admixture of the S-wave. 

In this paper we compare the prediction of
eq.~(\ref{ratiohard}) with the actual predictions of the NRQCD.  
From this angle it would be very important to look at the simplest process
of the coherent photoproduction of charmonia. This is naturally a point-like
process, in which the initial $c\bar c$ is produced in the S-state. Hence
for this process all deviations from eq.~(\ref{ratiohard}) would be due to
final state interactions. In particular it would be very interesting to
study the A-dependence of the $\psi''/\psi'$ ratio. 

Also for heavy ion physics this could be an interesting investigations. 
The ratio of eq.~(\ref{ratiohard}) will not be changed due to hadronic final
state interactions since these two states are very similar in size and mass.
This can be tested at CERN-SPS by the NA60 collaboration~\cite{na60} or at
RHIC by the PHENIX collaboration~\cite{phenix}.  Remember that for the
inclusive charm production the current data both at FNAL
energies~\cite{McGaughey:1999mq} and at RHIC~\cite{rhic} are consistent with
the linear dependence of the cross section on A which is expected within the
QCD factorization theorem for the kinematics of the discussed experiments. 

We discussed above that hadronic final state interactions will not change
the ratio of eq.~(\ref{ratiohard}). However, in heavy ion collisions the
formation of new phases is possible, e.g.\ the quark gluon plasma (QGP).  It
is not understood how to calculate this ratio in nonperturbative QCD in the
QGP. However, for heavy ion collisions the thermal production of charmonium
states is discussed in the literature~\cite{thermal}. In these models the
ratio of the $\psi''$ and the $\psi'$ yield is given by
\begin{equation}
{\sigma(\psi'')\over\sigma(\psi')}=
\left({m_{\psi''}\over m_{\psi'}}\right)^{3/2}\cdot
\exp\left({m_{\psi''}-m_{\psi'}\over T}\right)\quad.
\label{thermaleq}
\end{equation}
$T=170\div 180$~MeV in eq.~(\ref{thermaleq}) is the temperature, which is 
fitted to particle ratios in heavy ion collisions. Thus, 
eq.~(\ref{thermaleq}) 
yields in total
\begin{equation}
{\sigma(\psi'')\over\sigma(\psi')}=0.63\div 0.65\quad,
\end{equation}
which is much larger than ratio of the cross sections predicted for the hard
production processes. Moreover one should take into account that corrections
to eq.~(\ref{ratiohard}) due to the higher $\psi''$ mass decrease this ratio
even more. E.g.\ the pole mass of the propagators might yield a suppression
factor of $\left({m_{\psi'}\over m_{\psi''}}\right)^2= \left({3.686\over
3.77}\right)^2=0.96$. The suppression which comes due to the shift of the
initial partons to higher Bjorken x due to the higher mass yields for the
gluon distribution a factor $\left({1-3.686/\sqrt{s}\over
1-3.77/\sqrt{s}}\right)^{-12}$.  For the relatively low SPS-energies of
$\sqrt{s}=20$~GeV this suppression factor is 0.94. However, since these
corrections are smaller than 10$\%$ and since they even increase the
difference between soft and hard production we can safely neglect these
corrections here. Please note that for some experiments the kinematics can
be chosen in a way that the mass difference between the $\psi''$ and the
$\psi'$ becomes unimportant, e.g.\ in hadron-hadron and hadron-nucleus
collisions.  Also, the mass difference becomes unimportant at high
transverse momenta, since here the transverse mass not the rest mass of the
particles is the relevant parameter. 

To summarize, we have presented arguments in favour of a significant rate of
the $\psi''$ production in various hard processes and for the sensitivity of
the ratio of the $\psi''/\psi' $ rate to the dynamics of charmonium
production. A number of the current experiments can answer this question by
scanning their existing data.

\vspace*{1cm}
{\bf Acknowledgement:}\\
We thank E.~Braaten and T.~Ullrich for useful discussions. This work was
supported in part by GIF and DOE. L.G.\ thanks the Minerva Foundation for
support.


\begin{thebibliography}{99}

\bibitem{Bodwin:1994jh}
G.~T.~Bodwin, E.~Braaten and G.~P.~Lepage,
Phys.\ Rev.\ D {\bf 51}, 1125 (1995)
[Erratum-ibid.\ D {\bf 55}, 5853 (1997)]
[arXiv:hep-ph/9407339].

\bibitem{Bodwin:2002mr}
G.~T.~Bodwin,
arXiv:hep-ph/0212203.

\bibitem{babar}
B.~Aubert {\it et al.}  [BABAR Collaboration],
arXiv:hep-ex/0207097.

\bibitem{cleo}
S.~Anderson {\it et al.}  [CLEO Collaboration],
arXiv:hep-ex/0207059.


\bibitem{Eichten:1974af}
E.~Eichten, K.~Gottfried, T.~Kinoshita, J.~B.~Kogut, K.~D.~Lane and 
T.~M.~Yan,
Phys.\ Rev.\ Lett.\  {\bf 34} (1975) 369
[Erratum-ibid.\  {\bf 36} (1976) 1276].

\bibitem{Richard:1979fc}
J.~M.~Richard,
Z.\ Phys.\ C {\bf 4} (1980) 211.

\bibitem{FS91}
L.~Frankfurt and M.~Strikman,
Prog.\ Part.\ Nucl.\ Phys.\  {\bf 27}, 135 (1991).

\bibitem{cleo2}
G.~Bonvicini {\it et al.}  [CLEO Collaboration],
Phys.\ Rev.\ Lett.\  {\bf 84}, 5940 (2000)
[arXiv:hep-ex/0003004].

\bibitem{belle}
K.~Abe {\it et al.}  [BELLE Collaboration],
arXiv:hep-ex/0211047.

\bibitem{belle2}
K.~Abe {\it et al.}  [Belle Collaboration],
Phys.\ Rev.\ Lett.\  {\bf 89}, 142001 (2002)
[arXiv:hep-ex/0205104].

\bibitem{bodwin}
G.~T.~Bodwin, J.~i.~Lee and E.~Braaten,
arXiv:hep-ph/0212352.

\bibitem{cdf}
F.~Abe {\it et al.}  [CDF Collaboration],
Phys.\ Rev.\ Lett.\  {\bf 69}, 3704 (1992).

\bibitem{pdg}
K.~Hagiwara {\it et al.}  [Particle Data Group Collaboration],
Phys.\ Rev.\ D {\bf 66}, 010001 (2002).

\bibitem{e789}
M.~H.~Schub {\it et al.}  [E789 Collaboration],
Phys.\ Rev.\ D {\bf 52}, 1307 (1995)
[Erratum-ibid.\ D {\bf 53}, 570 (1996)].

\bibitem{e772}
D.~M.~Alde {\it et al.},
Phys.\ Rev.\ Lett.\  {\bf 66}, 133 (1991).

\bibitem{herab}
I.~Abt {\it et al.}  [HERA-B Collaboration],
arXiv:hep-ex/0211033.

\bibitem{nrqcd}
L.~K.~Hao, K.~Y.~Liu and K.~T.~Chao,
Phys.\ Lett.\ B {\bf 546}, 216 (2002)
[arXiv:hep-ph/0206226].\\
F.~Yuan, C.~F.~Qiao and K.~T.~Chao,
Phys.\ Rev.\ D {\bf 56}, 329 (1997)
[arXiv:hep-ph/9701250].\\
P.~w.~Ko, J.~Lee and H.~S.~Song,
Phys.\ Lett.\ B {\bf 395}, 107 (1997)
[arXiv:hep-ph/9701235].

\bibitem{na60}
B.~Lenkeit  [NA60-Collaboration],
arXiv:nucl-ex/0108015.

\bibitem{phenix}
K.~Adcox {\it et al.}  [PHENIX Collaboration],
Phys.\ Rev.\ Lett.\  {\bf 88}, 192303 (2002)
[arXiv:nucl-ex/0202002].

\bibitem{McGaughey:1999mq}
P.~L.~McGaughey, J.~M.~Moss and J.~C.~Peng,
Ann.\ Rev.\ Nucl.\ Part.\ Sci.\  {\bf 49} (1999) 217
[arXiv:hep-ph/9905409].

\bibitem{rhic}
D.~G.~D'Enterria  [PHENIX Collaboration],
arXiv:hep-ex/0205060.

\bibitem{thermal}
M.~I.~Gorenstein, A.~P.~Kostyuk, L.~McLerran, H.~St{\"o}cker and W.~Greiner,
J.\ Phys.\ G {\bf 28}, 2151 (2002).\\
P.~Braun-Munzinger and J.~Stachel,
Phys.\ Lett.\ B {\bf 490}, 196 (2000)
[arXiv:nucl-th/0007059].\\
R.~L.~Thews, M.~Schroedter and J.~Rafelski,
Phys.\ Rev.\ C {\bf 63}, 054905 (2001)
[arXiv:hep-ph/0007323].
\end{thebibliography}
\end{document}